# DECISION SUPPORT FACILITY FOR THE APS CONTROL SYSTEM*

D. A. Dohan, Advanced Photon Source, ANL, Argonne, IL 60561, USA

Abstract

The Advanced Photon Source is now in its fifth year of routine beam production. The EPICS-based [1] control system has entered the phase in its life cycle where new control algorithms must be implemented under increasingly stringent operational and reliability requirements. The sheer volume of the control system (~270,000 records, ~145 VME-based input-output controllers (IOCs), and ~7,000,000 lines of EPICS ASCII configuration code), presents a daunting challenge for code maintenance.

The present work describes a relational database that provides an integrated view of the interacting components of the entire APS control system, including the IOC low-level logic, the physical wiring documentation, and high-level client applications. The database is extracted (booted) from the same operational CVS repository as that used to load the active IOCs. It provides site-wide decision support facilities to inspect and trace control flow and to identify client (e.g., user interface) programs involved at any selected point in the front-end logic. The relational database forms a basis for generalized documentation of global control logic and its connection with both the physical I/O and with external high-level applications.

## 1 INTRODUCTION

The Advanced Photon Source facility has been in existence for about ten years, with the past five years under full beam production. The facility consists of an electron linac, followed by an accumulator ring and a synchrotron, injecting into an 1100-m storage ring at 7 GeV. There are typically between 20 and 30 parallel user experiments running at any time. After storing beam, the injector linac is used for free-electron laser development. Recently the storage ring has switched to top-up operation, in which beam is injected at 2-minute intervals, keeping the circulated beam current constant to within 1%. In order to allow FEL operation in parallel with top-up operation, an interleaving system is under development, allowing rapid switching of the linac beam between injection and FEL operation.

Concurrent with these ongoing requests for new beam capabilities and control features has been an increased emphasis on reliable operation, both in terms of reducing beam off time (MTTR) and perhaps more importantly, in terms of reducing the frequency of beam dropout (MTBF). Changes in the large installed, operational code base cannot be made without detailed understanding of the software and the behavior of the components of the control system.

The present paper describes work in progress at the APS to create a facility to assist the control systems application developer in analyzing the installed, operational control system. This decision support system utilizes a relational database approach to capture and query the controls knowledgebase.

## 2 THE APS CONTROL SOFTWARE

The main backbone of the APS accelerator control system [2] consists of a network of 145 front-end IOCs. Connected to the IOC layer is a network of operator interface workstations. Operators interact with the accelerator equipment mainly through MEDM [3] display files. Logging and automation tasks are handled using SDDS- and Tcl/TK-based tools. At the machine and equipment interface, embedded controllers are programmed to provide fixed well-defined functionality within limited control subdomains.

The subject of this paper is the middle, integration layer, which uses the EPICS real-time control system software.

## 3 PROCESS VARIABLE RELATIONAL DATABASE

### 3.1 The Relational Schema

EPICS software is based on the concept of 'records' – executable code units that perform hardware I/O. Other types of records (e.g., sequence, calc, fanout,) provide control logic capability. Each record has a well-defined set of 'fields,' commonly referred to as process variables or PVs. External clients may connect to these PVs, to read or optionally set their values. A particular type of field is a 'link' field, which is used to control the flow of logic between EPICS records. Records relating to a subsystem of the accelerator complex are grouped and stored in text files known as 'db files.' These db files assist the developer in organizing and partitioning software into manageable units of related control software. The IOC does not retain knowledge of the source db file from which any record is derived.

---

* Work supported by the U.S. Department of Energy, Office of Basic Energy Sciences under Contract No. W-31-109-ENG-38.

To provide an integrated view of the entire control system logic, a project was initiated to model the installed EPICS control software in a relational schema, using the ORACLE relational database (rdb). A simple schema was adopted, as shown in entity relation diagram of Figure 1.

Figure 1: EPICS PV entity relation diagram.

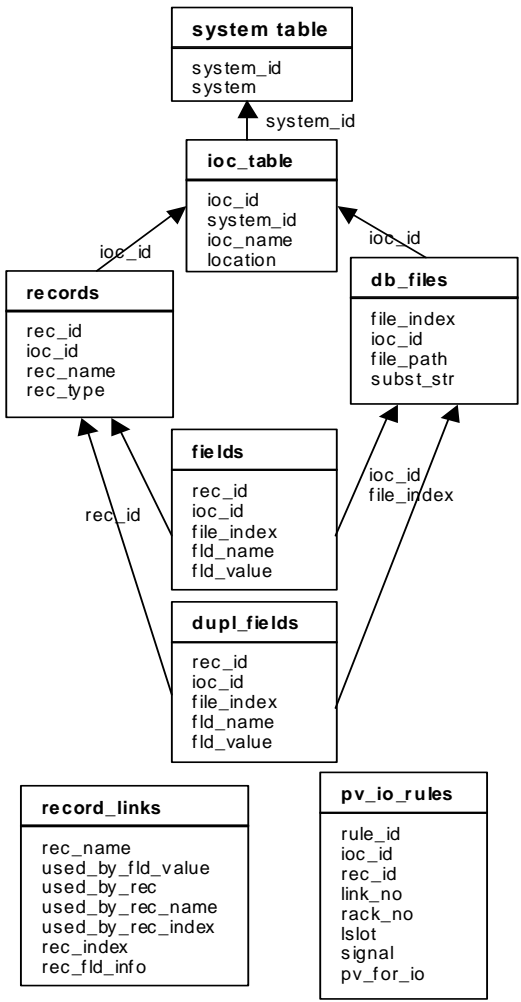

In this schema, the EPICS record and field names are treated as data rather than as attributes. Each of the fields defined in an EPICS db source file becomes a row in the fields table. In the APS control system, there are ~7,000,000 user defined fields. The foreign key for each entry in the fields table references a specific row in the records table. Each entry in the records table has a foreign key to a specific IOC. To assist in database drill down and access, the IOCs are grouped into systems organized by accelerator or by function.

## 3.2 Database Loading

The file descriptor of each IOC's startup command file is stored in its boot prom. At the APS, the standard boot procedure is modified such that this file descriptor is written out to an IOC-specific 'bootparams' file whenever an IOC reboots. A Perl script periodically checks the modification dates of these bootparams files to determine if any IOC has rebooted. On detection of a reboot, the startup file is parsed to determine the EPICS database definition and instance files to be loaded into the IOC. The EPICS static database access routines are used to extract the record and field definitions from the db files. The script writes this information, along with the respective db file paths in a set of flat files. The script determines if any fields have been multiply defined, and records this information along with their parent source filespecs. Although duplicate PVs are legal and ignored by the IOC, their usage presents a potential trap, which may be otherwise difficult to locate.

An ORACLE server was set up in a stand-alone workstation to contain the EPICS data. A periodic task in this server monitors the directories of the Perl extraction files, and triggers a loader process whenever new flat files have been created. The ORACLE tables are typically updated with the new IOC information within ~15 min after an IOC reboot. As a check of the validity of the ORACLE data, a nightly task is run in which flat files are generated from the ORACLE tables and compared with original flat files. This round trip file compare provides an independent check of the validity of the data.

## 3.3 User Queries – the pv_search Application

EPICS database inspection capabilities are provided using the standard ORACLE three-tier application server architecture. User requests, entered in a Web browser, are submitted to an ORACLE application server, which formats an SQL query and forwards it to the ORACLE server. Users enter optional IOC, db file, record name, record type, field name, and field type selection criteria. The user then selects one of the records returned by the query to obtain its field detail. In the detail display, EPICS link-type fields are displayed as hypertext links. Selecting a link field will display the details of the linked record.

The record_links table in Figure 1 is derived from the PV database and contains all instances of inter-record links in the entire set of IOCs. This derived table is used to extract the records linked to the actively displayed record. This list is displayed as hypertext links, allowing the user to extract the detail of the linking record. This powerful feature allows the user to bidirectionally explore EPICS logic.

*3.4 MEDM Clients*

At the APS, the primary operator interface to the accelerator is through the use of MEDM display files. These display files are stored in a single directory structure, under CVS control. A Perl script was written to examine all the files in this directory tree and to recursively extract references to EPICS process variables. When an EPICS record is displayed using the pv_search application, a list of all client MEDM screens that reference any field in the record is displayed. This identifies to the developer the display files that will need to be modified if, for example, the record name is changed.

*3.5 Decision Support - Data Mining*

The powerful querying capability of SQL has been used to 'data mine' the process variable database to provide a number of decision support queries.

**Orphan records**. This query extracts all records that:
- have no other PVs linked to it
- do not link to other PVs
- do not connect to hardware
- have no client application connection (MEDM, SDDS logger, etc.)

Records matching this category are potentially 'orphan' records, possibly left over from a prior or test subsystem. Since they probably provide no useful function, they are candidates for removal from the IOC.

**Multiply-defined hardware output records**. In the EPICS environment, it is legal to define more than one record for a single hardware I/O point. Although this feature is occasionally used as a convenience function, it can lead to equipment behavior problems if conflicting logic attempts to write to a single output channel. Queries have been written to extract and list instances of multiple hardware output records.

**Equipment database**. For EPICS records that are connected to external hardware I/O, the DTYP fields provide information related to the hardware device. This feature has been used to query the PV database to extract a list of hardware devices in the APS control system. The query report yields a total of 5800 individual devices directly associated with the APS control system. This equipment inventory is used in assisting a hardware spares strategy.

**Wiring List**. A project has been initiated to centralize wiring list documentation, using the Allen-Bradley equipment as a prototype. Allen-Bradley link, chassis, rack, slot, and terminal relational tables have been defined. A Web-based application allows developers to modify and update the ORACLE tables from any location that has access to a Web browser. Chassis and I/O card templates facilitate data entry. The derived pv_io_rules table shown in Figure 1 connects the wiring information with the process variable database. Display of the wiring details of an I/O card also lists the process variable connected to each terminal. This includes multiple PVs assigned to a single I/O point discussed above.

## 4 DISCUSSION

Complex and extensive mature systems like the APS control system have developed and evolved over years of field testing. Documentation of the rationale for coding decisions is often difficult to locate, or is incomplete or missing entirely. The control source code is often the only source of this documentation. A centralized control software inspection capability has proven to be useful in this effort. The PV-centric ORACLE relational database along with the querying power of SQL has provided intuitive mechanisms for locating relevant and related hardware and code documentation.

Work in progress includes extending the wiring database to include all the APS control hardware. This equipment 'connection' database will not only catalog all the hardware modules and ancillary equipment but will also indicate where each component is connected to the control system. It will be fully integrated with the PV database to allow cross queries between hardware and software.

Work is in progress to extend the PV clients to other types, including SNL, alarm handler, and SaveCompareRestore. The Perl extraction scripts are being expanded to parse the source files (db and startup files) for code documentation at the record, db file, and IOC level. The ultimate goal is to develop a central mechanism for creating and retrieving control system documentation.

## REFERENCES

[1] http://www.aps.anl.gov/epics has extensive information on all parts of EPICS.
[2] W. P. McDowell et al., "Status and Design of the Advanced Photon Source Control System," Proc. of the 1993 Particle Accelerator Conference, Washington, DC, May 1993, pp. 1960-62, 1993.
13] K. Evans Jr., "An Overview of MEDM," Proc. of ICALEPCS '99, Trieste, Italy, October 4–8, 1999, pp. 466-468, 2000.